\documentclass[12pt]{article}

\usepackage{graphicx}
\usepackage{amssymb}
\usepackage{color}
\usepackage{url}
\usepackage{hyperref}
\usepackage[affil-it]{authblk}
\usepackage[margin=1.2in]{geometry}

\newcommand{\bvec}[1]{{\mbox{{\boldmath$#1$}}}} 

\newcommand{\paren}[1]{{(#1)}}
\newcommand{\noparen}[1]{{#1}}
\newcommand{\eqnref}[1]{{\ref{#1}}}
\newcommand{\emdash}{{ --- }}
\newcommand{\newendash}{{ -- }}

\newcommand{\opencite}[1]{\cite{#1}}
\newcommand{\inlinecite}[1]{\cite{#1}}

\begin{document}

\title{Multi-Ridge Fitting for Ring-Diagram Helioseismology}
\author{Benjamin~J.~Greer
  \thanks{Electronic address: \texttt{benjamin.greer@colorado.edu}}}
\author{Bradley~W.~Hindman}
\author{Juri~Toomre}
  \affil{JILA and Department of Astrophysical and Planetary Sciences, University of Colorado, Boulder CO}

	\date{Dated: \today}

\maketitle

\begin{abstract}
Inferences of sub-surface flow velocities using local domain ring-diagram helioseismology depend on measuring the frequency splittings of oscillation modes seen in acoustic power spectra. 
Current methods for making these measurements utilize maximum-likelihood fitting techniques to match a model of modal power to the spectra. 
The model typically describes a single oscillation mode, and each mode in a given power spectrum is fit independently. 
We present a new method that produces measurements with greater reliability and accuracy by fitting multiple modes simultaneously. 
We demonstrate how this method permits measurements of sub-surface flows deeper into the Sun while providing higher uniformity in data coverage and velocity response closer to the limb of the solar disk. 
While the previous fitting method performs better for some measurements of low-phase-speed modes, we find this new method to be particularly useful for high phase-speed modes and small spatial areas.
\end{abstract}

\section{Introduction}
\label{sec:intro}

Helioseismology determines the structure and dynamics of the solar interior through analysis of seismic waves observed at the surface. 
Ring-diagram helioseismology (\opencite{hill_1988}; \opencite{basu_1999}; \opencite{haber_2002}) investigates sub-surface horizontal flows by measuring the direction-dependent frequency shift of oscillation modes. 
The Doppler shift of the frequency of an oscillation mode due to a sub-surface flow is expressed as
\begin{equation}
\delta \omega_{n}(\bvec{k}) = \bvec{k} \cdot \bvec{u}_{n}(k),
\label{eqn:freqshift}
\end{equation}
where $\bvec{k}$ and $n$ are the horizontal wavenumber and radial order of the oscillation mode, and $\bvec{u}_{n}(k)$ is a spatial average of the horizontal velocity within the Sun,
\begin{equation}
\bvec{u}_{n}(k) = \langle \bvec{v}(\bvec{r}) \rangle = \int \! K_{n}(\bvec{r};k) \ \bvec{v}(\bvec{r}) \ \mathrm{d}^{3}r.
\label{eqn:kerneldef}
\end{equation}
Here, $\bvec{v}(\bvec{r})$ is the true horizontal sub-surface flow velocity at any point \noparen{$\bvec{r}$} in the Sun, and $K_{n}(\bvec{r};k)$ is the weighting function\emdash or sensitivity kernel\emdash associated with each mode ($k$,$n$), which describes the spatial extent over which the true velocity is averaged to create a single frequency shift \cite{birch_2007}. 
The variation of the frequency shift as a function of direction (the frequency splitting) is measured to provide an estimate of $\bvec{u}_{n}(k)$.

While the interpretation of the frequency splitting is straightforward, the method of extracting it from the data is not. 
Traditionally, a model of the spectral power is fit to oscillation modes visible in power spectra of line-of-sight velocity observed in the photosphere. 
The model accounts for a frequency shift of the modal power as a function of horizontal direction, and this shift is directly translated into a velocity as shown in Equation \eqnref{eqn:freqshift}. 
The specifics of the fitting procedure determine how well the frequency splittings are measured, as well as what other qualities of the power spectra are taken into account. 
There are currently two commonly used fitting procedures in the \textit{Helioseismic and Magnetic Imager} (HMI) Ring-Diagram Pipeline \cite{bogart_2011a}. 
The first method considered in this article is one introduced by \inlinecite{haber_2002}, which fits a frequency-shifted Lorentzian model to individual modes. 
Since this method analyzes single ridges of modal power sequentially, we refer to it as the Single-Ridge Fitting method (SRF). 
The second method used in the HMI Pipeline, which will not be considered here, also fits modes independently \cite{basu_1999}, but uses a significantly different power model that includes asymmetries in modal power, two background terms, and a host of other wavenumber- and direction-dependent modifications.

In this article, we present a new fitting method that utilizes a model similar to that used in the SRF method, but modified to permit multiple radial orders to be fit simultaneously. 
The development of this Multi-Ridge Fitting (MRF) method is an attempt to improve upon the performance of the SRF method in terms of reliability and accuracy of measured frequency splittings. 
We present a comparison of the frequency splittings (interpreted as an average velocity) from a common data set processed with both fitting methods. 
We focus on the performance for measurements of modes that reach deepest into the Sun and for measurements made near the solar limb.
As metrics of reliability and accuracy we consider the fit success rate, the typical random errors in the measurements, and how well each method recovers a known velocity.

In Section \ref{sec:data} we outline the data-processing steps taken to prepare solar observations for helioseismic mode fitting. 
In Section \ref{sec:fitting} we describe the SRF and MRF fitting methods and how they differ in procedure.
In Section \ref{sec:comparison} we compare the performance of the two fitting methods using a common data set.
In Section \ref{sec:discussion} we discuss the implications of the results in the context of improving accuracy and data coverage in ring-diagram helioseismology.

\section{Data Acquisition}
\label{sec:data}

The observable we use to determine helioseismic frequencies is the line-of-sight Doppler velocity measured in the photosphere. 
The Doppler-velocity measurements are produced by HMI (\opencite{scherrer_2012}) aboard the Solar Dynamics Observatory (SDO). 
HMI captures full-disk Dopplergram images with a cadence of 45 seconds and an array size of 4096$\times$4096 pixels, resulting in a spatial resolution of roughly 350 km per pixel at disk center. 
Tiles of various sizes (\textit{e.g.}, $16^{\circ}$, $4^{\circ}$, $2^{\circ}$ in heliographic angle) are extracted from the full Dopplergram images with tile centers spaced evenly on a latitude\newendash longitude grid so that adjacent tiles overlap by half their diameter. 
The tile centers are shifted in longitude as time progresses at a rate corresponding to the measured surface differential rotation rate from \inlinecite{snodgrass_1984}. 
The purpose of this tracking is to minimize the signal of differential rotation and produce a more isotropic flow field. 
Each tile in the mosaic of tiles for a given tile size is tracked through 25.6 hours (2048 time steps), corresponding to about $15^{\circ}$ of solar rotation. 
In this article we refer to this tracking duration as one day or one realization.

The tiles are apodized in space and time and Fourier transformed to create a 3-D power spectrum for each tile. 
The spectra are transformed from Cartesian coordinates ($k_x$, $k_y$, $\nu$) to polar coordinates ($k$, $\theta$, $\nu$) so that oscillation modes at a constant wavenumber \noparen{$k$} can be easily extracted. 
The positive $x$-direction ($\theta=0$) is taken to be in the prograde zonal direction. 
The MRF code uses these spectra as input, while the SRF code requires an additional processing step. 
For the SRF method, the power is normalized by the average power computed at each wavenumber \noparen{$k$} and azimuth angle \noparen{$\theta$} \cite{haber_2000}. 
The purpose of this filtering is to eliminate large power variations as a function of azimuth which arise from a variety of sources including camera astigmatism and power foreshortening. 
The MRF method does not use filtered power spectra, as it allows for such variations of power in the fitting procedure itself. 


\section{Fitting Methods}
\label{sec:fitting}

\subsection{Single-Ridge Fitting (SRF)}
\label{sec:srf}

\begin{figure}
	\centering
	\includegraphics{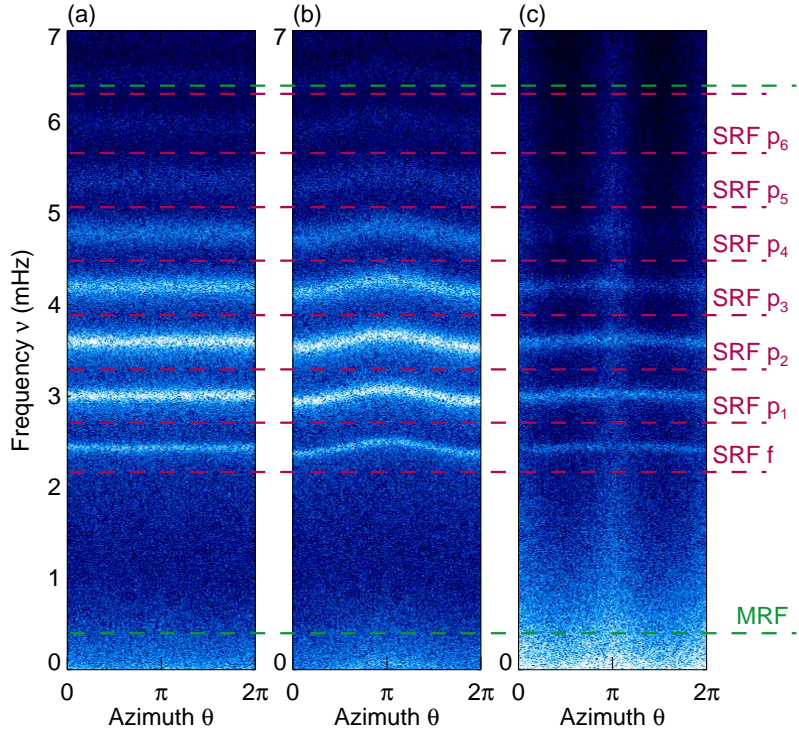}
	\caption{Sections at $k=0.8 \ \mathrm{Mm^{-1}}$ of the power spectra for three different $16^{\circ}$ tiles: 
	(a) tile extracted from disk center tracked at the Snodgrass rate, 
	(b) tile from disk center tracked at $500 \ \mathrm{m \ s^{-1}}$ relative to the Snodgrass rate, and 
	(c) tile from the central meridian and $75^{\circ}$N latitude tracked at $130 \ \mathrm{m \ s^{-1}}$.
	Red dashed lines indicate approximate bounds for the frequency windows used for independent mode fitting in the SRF method. 
	Green dashed lines indicate the single frequency window used in the MRF method. 
	Velocity-induced frequency shifts are evident in (b) and (c), and significant power foreshortening is visible in (c).}
	\label{fig:pspec2}
\end{figure}

In a cylindrical section at constant wavenumber \noparen{$k$}, distinct modes can clearly be seen (Figure \ref{fig:pspec2}). 
Each radial order intersects the surface of constant wavenumber to create a band of power with a central frequency. 
Sub-surface flows cause each band to trace out a slight sinusoidal undulation in frequency \paren{$\nu$} as a function of azimuth \paren{$\theta$}. 
In order to separate each radial order seen in the data, the SRF method extracts a small range of frequencies at a constant wavenumber centered around a specific mode. 
The bounds in frequency are based on a guess table that provides the initial values of the parameters that will be fit to the data. 
When a mode is framed both above and below in frequency by other modes, the bounds in frequency are taken as part-way between the central frequencies of the adjacent modes. 
For the highest and lowest modes in frequency that do not have adjacent modes on both sides, the outer boundaries are taken as a few linewidths away. 
SRF sequentially extracts frequency windows around each desired radial order and fits a six-parameter model (Equation \eqnref{eqn:slf}) to each window using a maximum-likelihood technique (\opencite{anderson_1990}; \opencite{haber_2002}). 
The SRF method represents the power in each mode as a symmetric Lorentzian with an angle-dependent frequency:

\begin{equation}
P(\nu, \theta) = B + \frac{A \Gamma/2}{[\nu - \nu_0 + (2 \pi)^{-1} \bvec{k} \cdot \bvec{u}]^2 + (\Gamma/2)^2},
\label{eqn:slf}
\end{equation}

\begin{equation}
\bvec{u} = u_x \ \mathrm{cos} \theta \ \bvec{\hat{x}} + u_y \ \mathrm{sin} \theta \ \bvec{\hat{y}}.
\end{equation}
Here $P(\nu, \theta)$ is the power as a function of frequency and azimuth; $A$ is the amplitude; $\Gamma$ is the line-width; $\nu_0$ is the central frequency; $k$ is the wavenumber; $u_x$ and $u_y$ are the Doppler velocity components; and $B$ is a constant background. 
There are six parameters in this model \paren{$A$, $\Gamma$, $\nu_0$, $u_x$, $u_y$, $B$} that need to be fit to each mode independently at every wavenumber.
By marching through each discrete wavenumber for a given tile size and attempting to fit each mode listed in the guess table, the SRF method produces the optimal values of these six parameters for every mode.
A measurement of the random error of each velocity component \paren{$u_x$ and $u_y$} is determined by the curvature of the maximum-likelihood function evaluated at the point of optimization \cite{anderson_1990}. 
It is this that we identify as the error or uncertainty in each measurement that we will be making.

There are a number of ways in which a fit can fail or be deemed invalid. 
The numerical optimization procedure can fail to converge on a solution, causing no valid data to be produced for a single mode. 
If there is a successful fit, the data can still be rejected if the parameters are outside of predetermined bounds. 
These bounds are chosen to ensure that the parameters obtained through fitting are physically relevant. 

\begin{figure}
	\centering
	\includegraphics{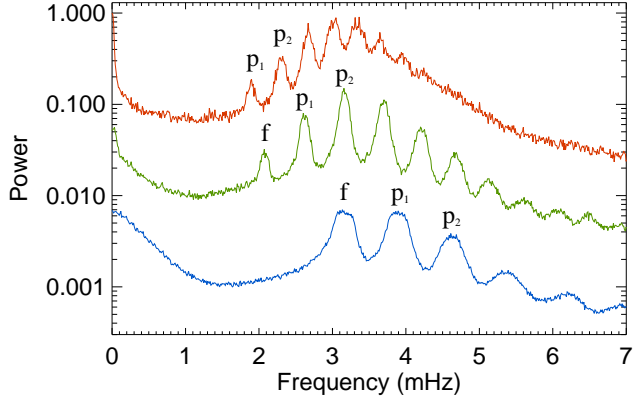}
	\caption{Azimuthal averages of a spectrum generated from a single $16^{\circ}$ tile extracted at disk center. Each curve corresponds to a different horizontal wavenumber: $k=0.23 \ \mathrm{Mm^{-1}}$ (red, top), $k=0.58 \ \mathrm{Mm^{-1}}$ (green, middle), and $k=1.42 \ \mathrm{Mm^{-1}}$ (blue, bottom). 
	Curves are offset in power for clarity. 
	Modes at lower wavenumber and higher frequency are spaced closer together, causing significant overlap. 
	Ridges at high wavenumber look flattened due to azimuthal averaging over a flow-induced frequency splitting. }
	\label{fig:blending}
\end{figure}

At moderate wavenumbers \paren{$k\ge0.5 \ \mathrm{Mm^{-1}}$} in larger tile sizes, individual modes are sufficiently separated in frequency to justify this approach of  independent fitting (Figure \ref{fig:blending}). 
However, at lower wavenumbers, modes of neighboring radial order blend together to form a mound of spectral power without prominent individual peaks. 
Smaller tile sizes also have more blending of modes at all wavenumbers due to the smaller spatial apodization. 
In these situations, severe mode blending brings into question the idea of fitting each mode independently with a single Lorentzian model with a constant background term. 
The frequency window used around each mode also raises issues when a sub-surface flow introduces a large frequency splitting. 
At $k=0.5 \ \mathrm{Mm^{-1}}$, a $300 \ \mathrm{m \ s^{-1}}$ flow is sufficient to shift significant mode power past the frequency boundaries chosen to split up the data for fitting. 
Tiles produced in the standard HMI Ring-Diagram Pipeline are tracked at the Carrington rate \cite{bogart_2011a}, which at extreme latitudes differs from the local surface rotation rate by nearly $250 \ \mathrm{m \ s^{-1}}$.
Not only does mode power shift outside the fitting window for a given mode, but power from neighboring modes that is not accounted for in the model enters the window and potentially throws off the fitting \cite{routh_2011}. 

\subsection{Multi-Ridge Fitting (MRF)}
\label{sec:mrf}

To mitigate the issues that come with independent fitting of modes, we have developed the Multi-Ridge Fitting (MRF) code.
This new method fits multiple radial orders simultaneously at a given wavenumber over a wider window in frequency. 
The function used to describe the collection of modes is a sum of symmetric Lorentzians, each similar to those used in the SRF method:
\begin{equation}
P(\nu, \theta) = B(\nu,\theta) + \sum\limits_{n=0}^{N-1} \frac{A_n (\Gamma_n / 2)}{[\nu - \nu_n + (2 \pi)^{-1} \bvec{k} \cdot \bvec{u}_n]^2 + (\Gamma_n/2)^2} F(\theta;f_n,\theta_n),
\label{eqn:mrf}
\end{equation}
\begin{equation}
F(\theta;f_n,\theta_n) = 1 + f_n \mathrm{cos} [2(\theta - \theta_n)],
\label{eqn:mrf2}
\end{equation}
\noindent
where $A_n$, $\Gamma_n$, $\nu_n$, $u_{x,n}$, and $u_{y,n}$ retain the same purpose as in the SRF method,  
and the sum is taken over the number of modes specified in the guess table at a given wavenumber. 
The new factor \noparen{$F(\theta;f_n,\theta_n)$} in the numerator of each Lorentzian accounts for amplitude variation as a function of $\theta$ using the parameters $f_n$ and $\theta_n$. 
While the constant background term used in Equation \eqnref{eqn:slf} is roughly valid when considering the narrow range of frequencies used in SRF, the background term is modified here to represent background power over a wider range by using a modified Harvey law \cite{harvey_1985}:
\begin{equation}
B(\nu, \theta) = \frac{B_0}{1+(\nu/\nu_{\mathrm{bg}})^b} F(\theta;f_\mathrm{bg},\theta_\mathrm{bg}).
\label{eqn:mrf_back}
\end{equation}
Here, $B_0$ is the amplitude; $\nu_{\mathrm{bg}}$ is a roll-off frequency; and $b$ is the power law index. 
Again, there is a power anisotropy term \paren{$f_{\mathrm{bg}}$, $\theta_{\mathrm{bg}}$} in the amplitude. 
The low-frequency end of the fitting window is taken as $0.3 \ \mathrm{mHz}$ at all wavenumbers to allow a large fraction of the background power to be utilized in constraining the model. 
The top of the frequency window is taken as two linewidths above the highest central frequency listed in the guess table at a given wavenumber. 
As a pre-conditioning step before performing the optimization of every parameter in Equation \eqnref{eqn:mrf}, the MRF method performs a simple three-parameter fit \paren{$A$, $\Gamma$, $\nu_0$} to each mode listed in the guess table. 
This step is similar to the fit performed in the SRF method, but does not take any variation along azimuth into account. 
This step provides an improved initial condition for the full optimization for each mode. 
Improved initial parameters for the background term are also obtained prior to the final optimization by fitting the parameters ($B_0$, $\nu_{\mathrm{bg}}$, $b$) to the spectrum between $0.3 \ \mathrm{mHz}$ and $1.5 \ \mathrm{mHz}$. 
Random-error estimates are once again measured from the curvature of the maximum-likelihood function.

It is important to note the significant increase in the number of parameters from Equation \eqnref{eqn:slf} to Equation \eqnref{eqn:mrf}. 
Typical nonlinear optimization algorithms scale in time as $m^2$, where $m$ is the number of parameters being fit simultaneously. 
To fit all of the ridges at a single wavenumber, the SRF method performs $N$ sequential optimizations with $m=6$. 
The MRF method fits the same set of ridges simultaneously with $m=7N+5$, leading to an expected optimization time $\approx1.4N$ times slower for large $N$. 
Not only is this new procedure more computationally demanding, it is much more susceptible to the numerical issue of wandering through parameter space. 
To ensure a timely convergence, constraints are placed on some of the model parameters:
\begin{itemize}
\item Amplitudes and widths must be positive.
\item Central frequencies must not cross those of adjacent modes.
\item Widths must be within an order of magnitude of lookup table values.
\item Both zonal and meridional velocities must be smaller than $\pm$ $\mathrm{1 km \ s^{-1}}$.
\item Fractional anisotropy must be between 0 and 1.
\end{itemize}
These constraints are enforced during the optimization process to limit the available parameter space.

\section{Comparison}
\label{sec:comparison}

\begin{figure}[]
	\centering
	\includegraphics{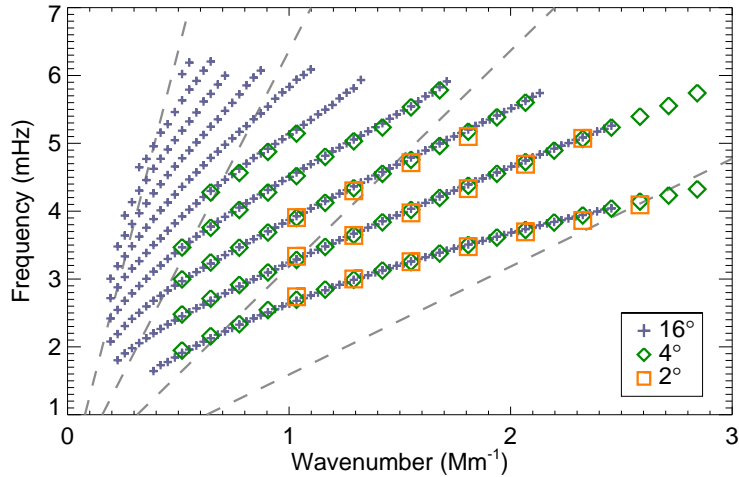}
	\caption{Guess-table frequencies for each tile size demonstrating which modes are attempted in both fitting methods. 
	The gray dashed lines are lines of constant phase-speed \paren{$\omega/k$}. 
	From left to right, the values of phase-speed are $10$, $20$, $40$, and $80 \ \mathrm{km \ s^{-1}}$. 
	The discrete ridges of modes indicate the individual radial orders, with the $f$-mode at low frequency, followed by $p_1$ through $p_{9}$.
	}
	\label{fig:modeset}
\end{figure}

Each fitting method requires a set of initial values for each parameter, often referred to as a guess table.
The guess table contains typical values of the central frequencies, widths, and amplitudes of each mode that are fit. 
To have a fair comparison between the two methods, they have been supplied the same guess table, and thus the same set of modes to attempt for each tile size (Figure \ref{fig:modeset}).

While the MRF method accounts for power anisotropy in the model (Equation \eqnref{eqn:mrf}), the SRF method depends on additional processing of the input power spectra to flatten power anisotropy. 
The unfiltered power spectra are fit using both methods, while the filtered ones are only fit by the SRF method. 
We confirm that the SRF method performs slightly better on the filtered spectra (Haber, personal communication, 2013), so the analysis in this article compares unfiltered spectra fit with the MRF method to filtered spectra fit with the SRF method.


\subsection{High Phase-Speed Modes}
\label{sec:deep}

Each distinct mode that is fit has an associated kernel that specifies the spatial sensitivity to sub-surface flows (see Equation \eqnref{eqn:kerneldef}). 
While the horizontal profile of these kernels is largely determined by the tile apodization (\opencite{hindman_2005}; \opencite{birch_2007}), the vertical profile changes significantly across the mode set as it depends on the mode's radial eigenfunction. 
The sensitivity for $f$-mode kernels has a single peak in depth, and each successively higher radial order adds an additional peak. 
The final peak of sensitivity for any mode occurs approximately where the horizontal phase-speed \noparen{$\omega/k$} equals the local sound speed, causing each mode within a single radial order but different horizontal wavenumber to have a slightly different kernel. 
By matching phase-speeds to the sound speeds tabulated in the tabulated solar Model S \cite{dalsgaard_1996}, we can use phase-speed as a proxy for the depth to which a mode reaches into the Sun. 

Current results that utilize the SRF method can reach down to 20 Mm (0.97 $R_{\odot}$), partway into the near-surface shear layer \cite{haber_2000}.
In order to extend modern ring-diagram analysis deeper into the Sun, we must obtain frequency-splitting measurements for higher phase-speed modes. 
An easy way to do this is to increase the tile size or the tracking duration, with the result of increasing the signal-to-noise across the entire power spectrum of each tile. 
Increased tile sizes also suffer less blending and can sample lower wavenumbers. 
However, if we wish to preserve horizontal and temporal resolution, we must consider how improvements to the fitting method can produce reliable measurements for higher phase-speed modes of a constant tile size. 

The first aspect to consider is how reliable each method is for obtaining measurements of high phase-speed modes. 
By dividing the number of successful measurements by the number attempted, we get a measure for each mode of the success rate on a per-mode basis. 
There are two primary reasons that either fitting method will fail to produce a successful measurement: poor data quality and poor choice of model. 
In the first case, as the signal-to-noise ratio of a particular mode decreases, the measured error on each model parameter increases. 
At some point, the fitting methods will fail to locate the mode amongst the noise and the fit will be deemed a failure. 
In the second case, if there is no way for the fitting method to adjust the provided model in a way that appropriately matches the data, it will struggle to converge on an optimal and unique solution. 
Both fitting methods will judge the attempt a failure if it does not converge within a specified number of optimization iterations or converges to non-physical model parameters. 
Measurements of the error on the model parameters for this case are not necessarily an accurate estimation due to the irreparable disparity between model and data.

\begin{figure}
	\vspace{0.03\textwidth}
	\centerline{
		\hspace*{0.015\textwidth}
		\includegraphics[clip=]{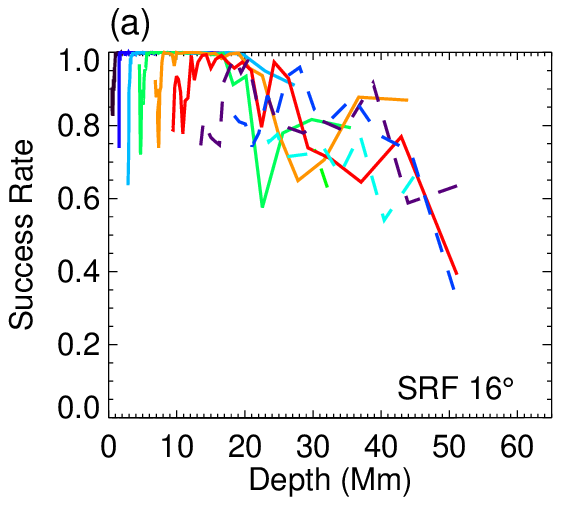}
		\hspace*{-0.03\textwidth}
		\includegraphics[clip=]{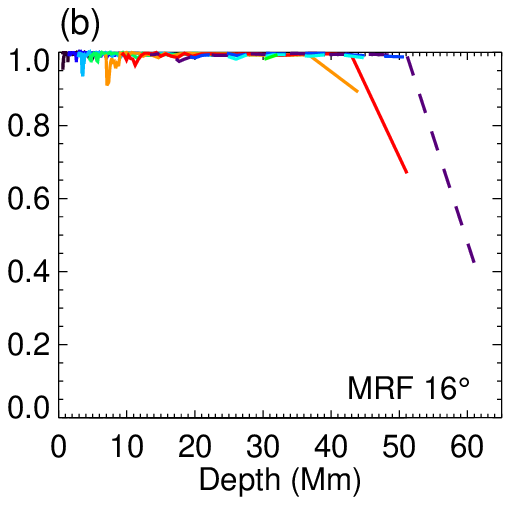}
	}
	\vspace{0.01\textwidth}
	\centerline{
		\hspace*{0.015\textwidth}
		\includegraphics[clip=]{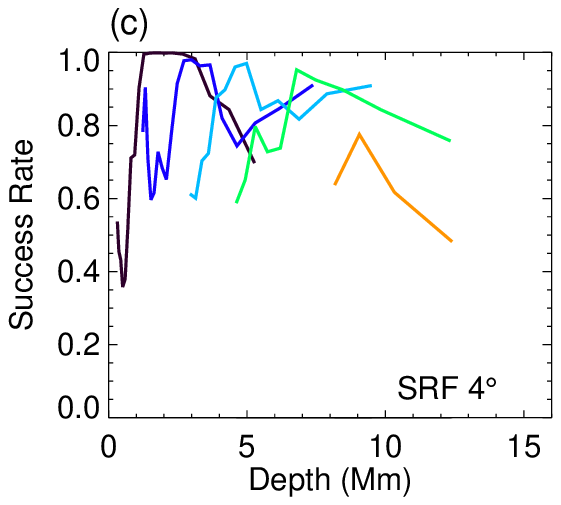}
		\hspace*{-0.03\textwidth}
		\includegraphics[clip=]{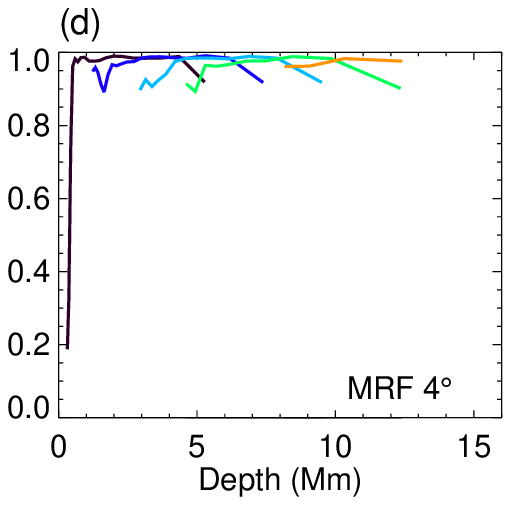}
	}
	\vspace{0.01\textwidth}
	\centerline{
		\hspace*{0.015\textwidth}
		\includegraphics[clip=]{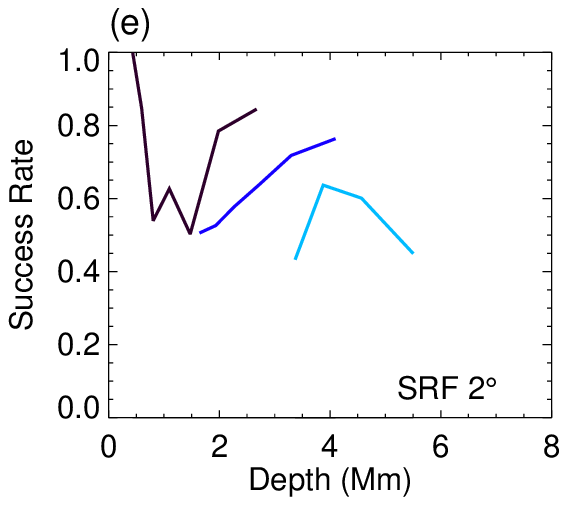}
		\hspace*{-0.03\textwidth}
		\includegraphics[clip=]{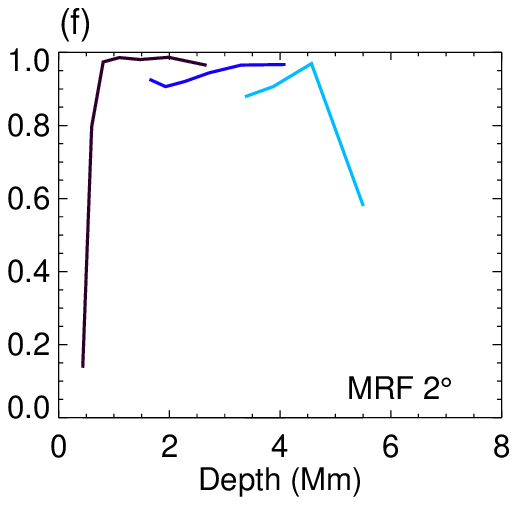}
	}
	\vspace{0.01\textwidth}
	\centerline{
		\hspace{0.05\textwidth}
		\includegraphics[clip=]{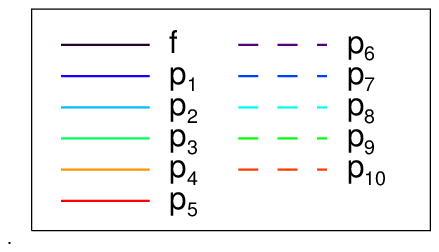}
	}
\caption{Fit success rate as a function of lower turning point of each mode found in the guess table. Left column is SRF results. 
Right column is MRF results. 
Rows correspond to the three tile sizes. 
To minimize the effects of disk position, all tiles of a given size falling within $30^{\circ}$ of disk center have been considered here. 
The success rate is averaged over ten independent realizations of these tiles, covering ten days of tracking. 
The MRF success rate is higher at nearly every depth for every tile size.}
\label{fig:successwk}
\end{figure}

Plotted as a function of lower turning point depth, Figure \ref{fig:successwk} demonstrates how deep into the Sun each fitting method is capable of measuring. 
To isolate the effects of phase-speed, the success rate illustrated here was compiled using only tiles bounded within $30^{\circ}$ of disk center.
Since the spacing between adjacent tiles in our mosaic depends on the tile size considered, there is a different number of tiles within this boundary for each tile size (49 for $16^{\circ}$ tiles, 805 for $4^{\circ}$, and 3249 for $2^{\circ}$).
We average the success rate over ten independent realizations of all tiles found within this boundary, resulting in 490 attempted velocity measurements of each mode for $16^{\circ}$ tiles, 8050 for $4^{\circ}$ tiles, and $\mathrm{32\,490}$ for $2^{\circ}$ tiles. 
While the guess table has been tuned to provide optimal results for the SRF method, the scatter seen in the SRF success rate is primarily caused by imperfect guess parameters. 
The MRF method is in general less sucsceptible to this effect. 
A striking difference between the two methods is how the success rate falls off with increasing depth. 
Both methods are most successful for the shallowest depths sampled with a given tile size. 
The SRF method tends to have a gradual decrease in the success rate as the depth increases, while the MRF method maintains a nearly constant success rate through the entire mode set. 
As the tile size decreases, the MRF success rate again stays largely constant while the SRF success rate decreases for all depths. 
The success rate for the SRF method also has a slight dependence on wavenumber, causing the variation of success rate between each radial order at every depth.

\begin{figure}
	\centering
	\includegraphics{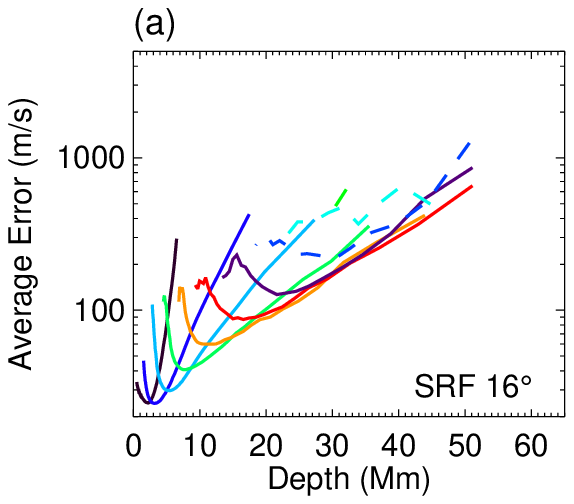}
	\includegraphics{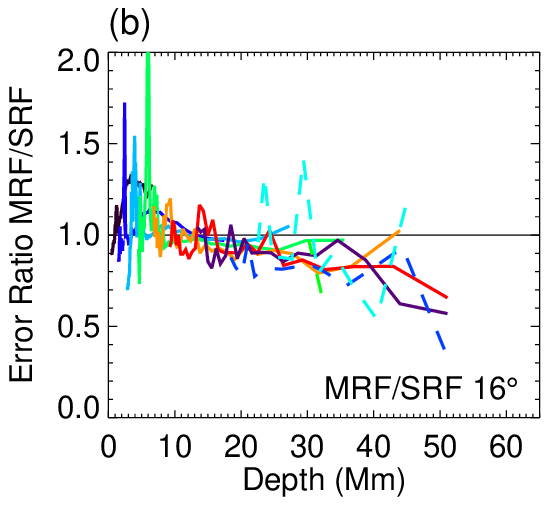}
	\includegraphics{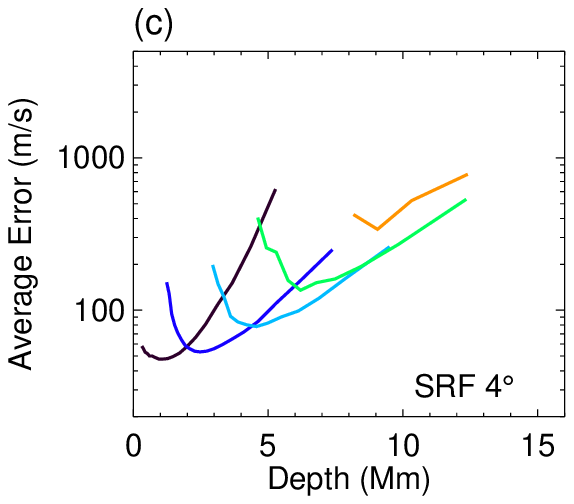}
	\includegraphics{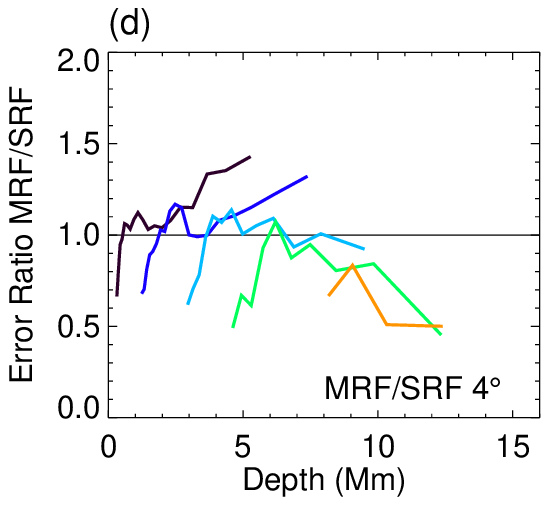}
	\includegraphics{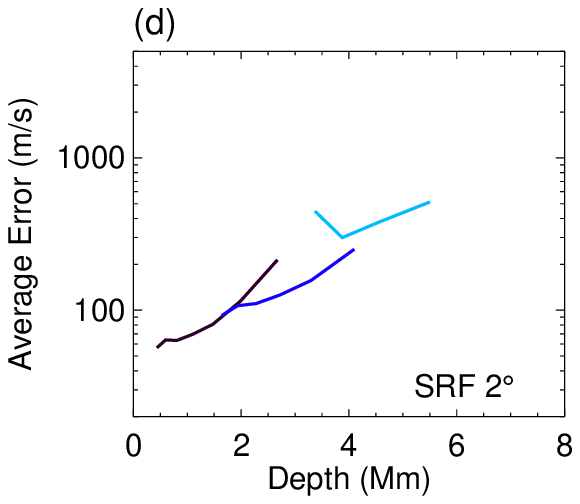}
	\includegraphics{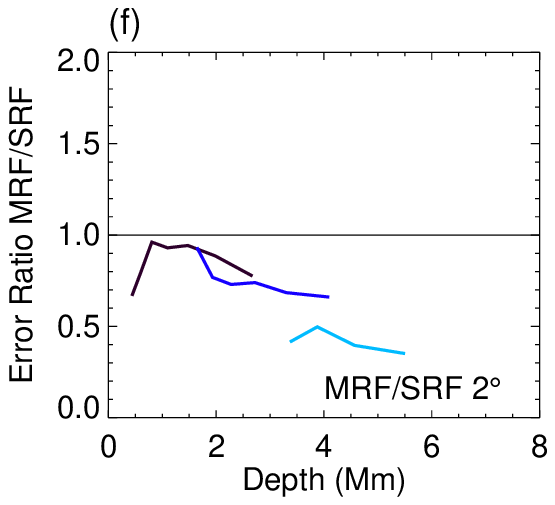}
\caption{Left: Average random error of the SRF method as a function of lower turning point for each mode. Right: Ratio of MRF errors to SRF errors. 
Rows correspond to the three tile sizes. 
To minimize the effects of disk position, all tiles of a given size falling within $30^{\circ}$ of disk center have been considered here.
The success rate is averaged over ten independent realizations of these tiles, covering ten days of tracking.
The MRF method produces errors that are largely similar to those from the SRF method, with some differences for modes that reach deepest. }
	\label{fig:avgerr}
\end{figure}

The measured random errors as determined by each fitting method also show a strong dependence on phase-speed (Figure \ref{fig:avgerr}). 
As a function of lower turning-point depth for each mode, the average random error is bounded from below by an envelope that exponentially increases with depth. 
Near-surface flows in the upper few megameters of the Sun (typical flow speed $\approx150 \ \mathrm{m \ s^{-1}}$) can be determined with sufficient precision using a single mosaic of tiles tracked through one day, while flows at the bottom of the near surface shear layer ($\approx35 \ \mathrm{Mm}$, typical flow speed $\approx50 \ \mathrm{m/s}$: \opencite{schou_1998}) require a significant amount of averaging in time and space to achieve a reasonable signal-to-noise ratio. 
Despite the considerable differences in success rate at extreme depths, there are only slight differences between the typical measured errors of each method. 
The average error obtained with the MRF method is slightly larger for shallow modes than that obtained with the SRF method. 
For deeper penetrating modes and for smaller tile sizes, the MRF errors become smaller than the SRF errors.

Through these two metrics (success rate, measured errors), we see that the MRF method performs more consistently with depth while maintaining similar estimates of the average error. 

\subsection{Approaching the Solar Limb}
\label{sec:limb}
The previous results regarding how each fitting method performs at high phase-speeds were compiled using tiles near the center of the solar disk \paren{$r\le30^{\circ}$}. 
Both methods perform optimally near disk center and produce lower-quality results closer to the limb. 
The primary cause of this is foreshortening, which causes the effective resolution in one direction to be considerably less than that in the perpendicular direction. 
In the power spectra, this appears as a reduction of power along the direction of foreshortening.
Since the MRF method has additional terms to account for this type of power variation in both the mode power and background power, it is expected that it will outperform the SRF method near the limb. 

The ability to push closer to the limb has many benefits. 
Regions of strong magnetic activity are continually evolving as they pass across the solar disk from the east limb to the west limb. 
To maximize the amount of time that these regions are accessible with ring-diagram analysis, we must be able to measure flow velocities close to the solar limb. 
The consistency of velocity measurements as a function of disk position is important for studying the evolution of flows across the disk. 
The possibility of high-latitude meridional counter-cells has been of great interest (\opencite{upton_2012}; \opencite{komm_2013}; \opencite{gonzalezhernandez_2010}), increasing the need for analysis methods that perform consistently between disk center and high latitudes. 
Careful removal of large-scale systematics has a noticeable effect on the determination of sub-surface flows (\opencite{zhao_2012}, \opencite{greer_2013}, \opencite{zhao_2013}). 
The accurate measurement of these systematics is also dependent on the spatial uniformity of the methods that obtain velocity measurements. 

\begin{figure}[]
	\centering
	\includegraphics{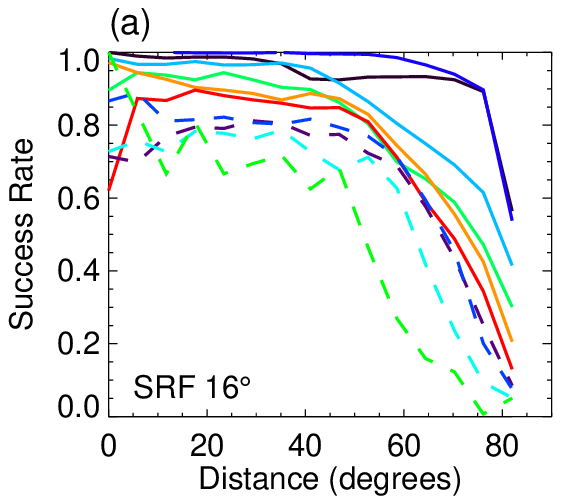}
	\includegraphics{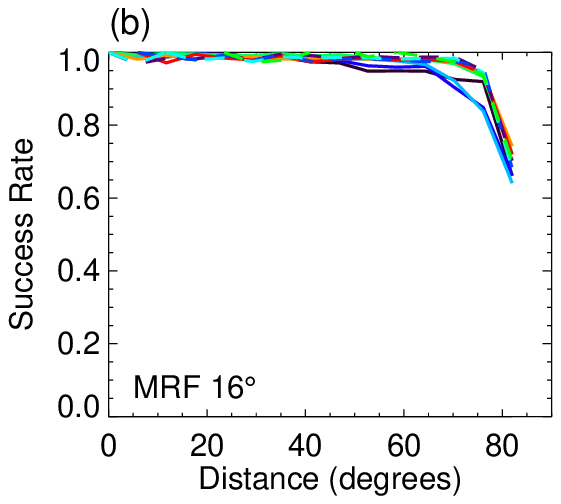}
	\includegraphics{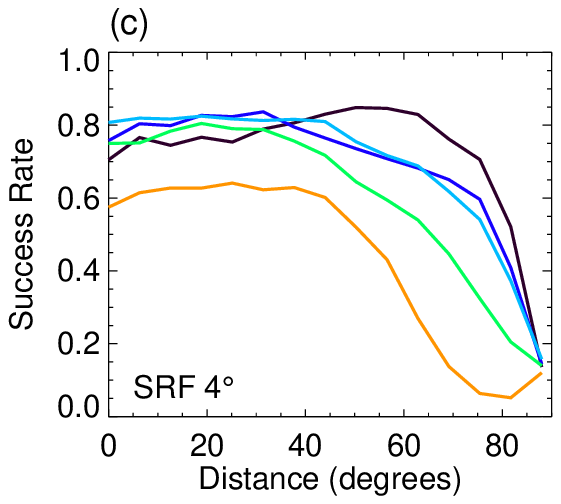}
	\includegraphics{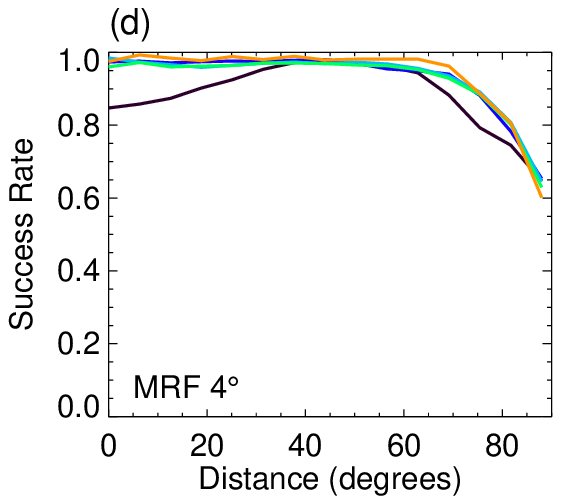}
	\includegraphics{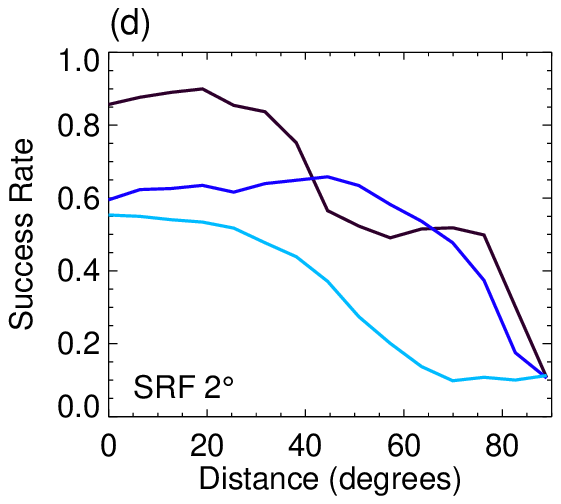}
	\includegraphics{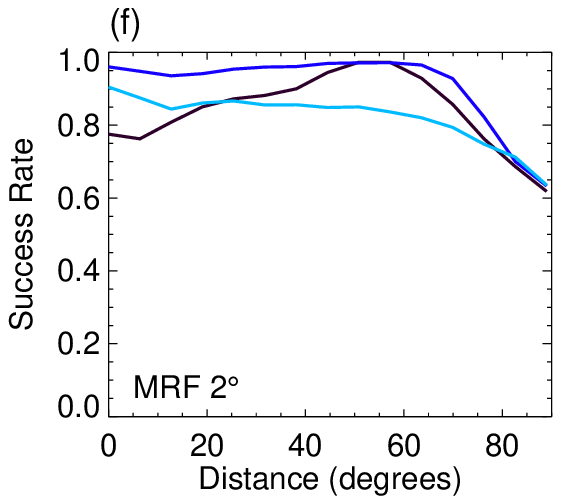}
	\includegraphics{legend.ps}
	\caption{Success rate for each radial order averaged over all wavenumbers as a function of distance from disk center. 
	The success rate as a function of distance from disk center for the SRF method is different for each radial order, while the success rate for the MRF method is largely constant. 
	The f-mode has roughly the highest success rate, while each higher radial order has a successively lower success rate.}
	\label{fig:successdisk}
\end{figure}

Once again, it is useful to first consider the success rate of each method as a function of disk position (Figure \ref{fig:successdisk}). 
A single mosaic of tiles tracked through one day covering the entire solar disk was analyzed to produce $\mathrm{223\,750}$ attempted measurements for $16^{\circ}$ tiles, $\mathrm{611\,390}$ for $4^{\circ}$ tiles, and $\mathrm{632\,740}$ for $2^{\circ}$ tiles.
For the SRF method, not only does the success rate at every disk position depend on the radial order of a mode, but the success rate for each radial order also falls off at different distances from disk center.
Each radial order has a distance from disk center where the success rate transitions from a nearly constant value near disk center ($0^{\circ}$) to a steep drop towards the limb. 
For $16^{\circ}$-tile $f$-mode fits made with the SRF method, this distance is around $75^{\circ}$ in heliographic angle, but for each higher order the distance decreases. 
By $p_9$, this distance has dropped to only $45^{\circ}$. 
With fewer high phase-speed measurements being made away from disk center, the depth to which sub-surface flows can be determined becomes position dependent. 
The smaller tile sizes display a similar trend, only with a lower success rate for each order at all disk positions. 

In contrast, the MRF method has a higher success rate in nearly all cases. 
For $16^{\circ}$ tiles, the distances at which each radial order begins to drop off are now nearly equal at around $75^{\circ}$. 
This results in the MRF method producing much more uniform results across the solar disk, both in spatial coverage and in depth. 
Smaller tile sizes show a decrease in the success rate closer to disk center, but it is still roughly constant for all modes. 

\begin{figure}[]
	\centering
	\includegraphics{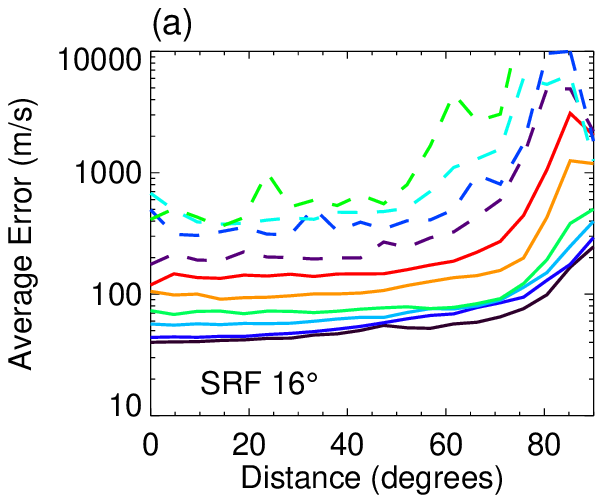}
	\includegraphics{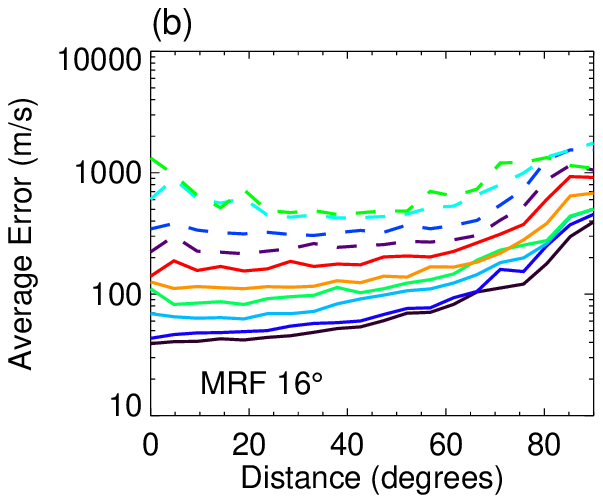}
	\includegraphics{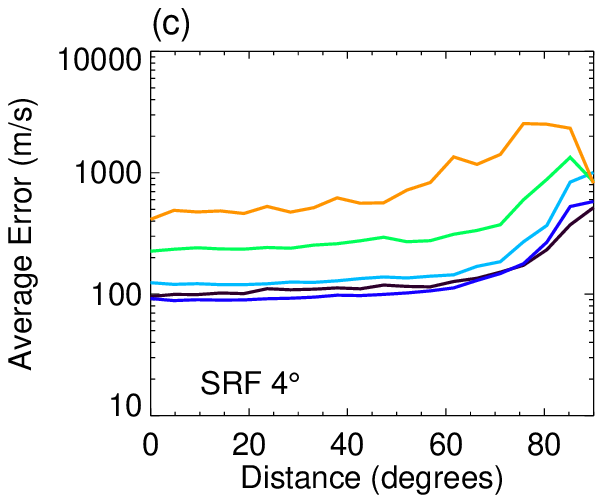}
	\includegraphics{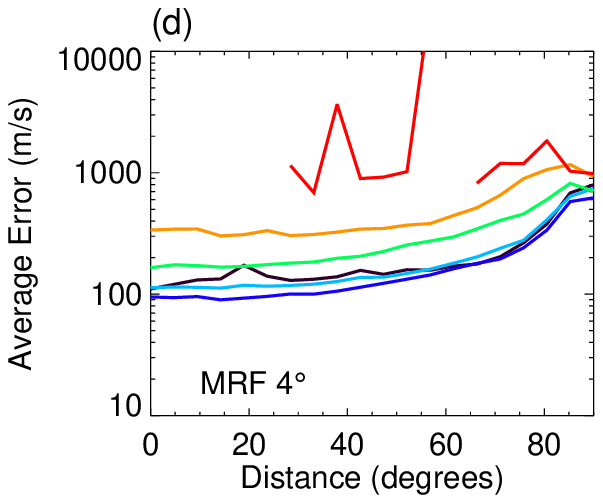}
	\includegraphics{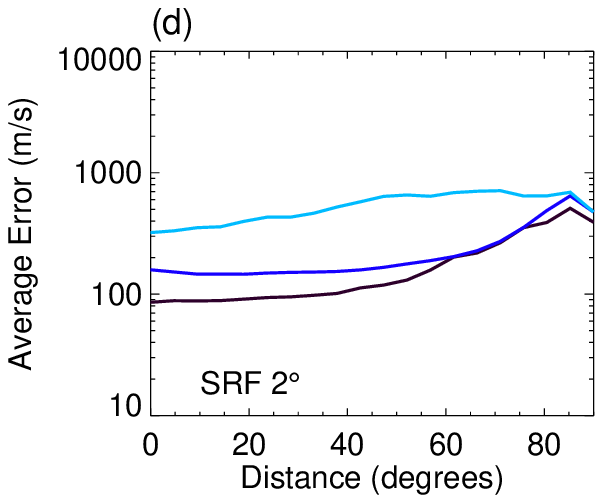}
	\includegraphics{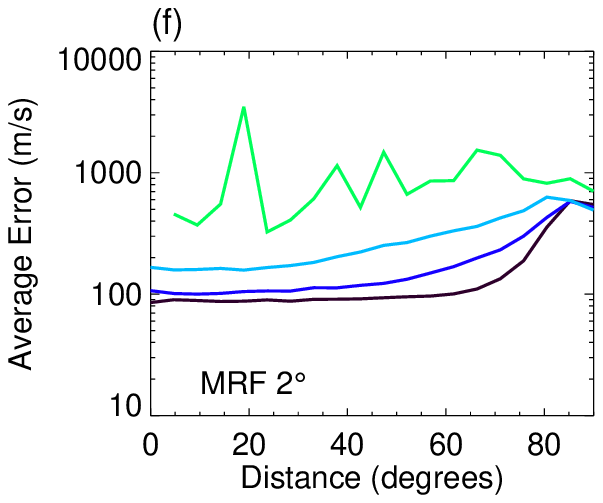}
	\caption{Average random error as a function of distance from disk center. 
		The average error for both methods stay roughly constant near disk center and increase dramatically outside $60^{\circ}$.
		The f-mode has the lowest average error, while each higher radial order has a successively higher average error.}
	\label{fig:errordisk}
\end{figure}

The average error as a function of distance from disk center (Figure \ref{fig:errordisk}) shows similar trends for both fitting methods. 
Both have nearly constant errors from disk center out to around $60^{\circ}$. 
Beyond this, the magnitude of the errors for all radial orders rise steeply in both fitting methods. 
The primary difference between the two methods is how severe this rise in error is near the limb. 
For $16^{\circ}$ tiles fit with the SRF method, the average error for high-radial-order modes quickly passes $1 \ \mathrm{km \ s^{-1}}$ outside of $60^{\circ}$ from disk center. 
The MRF method shows a slower rise in error for these modes, reaching $1 \ \mathrm{km \ s^{-1}}$ only for the worst cases.

\subsection{Velocity Accuracy}
\label{sec:velocities}

As demonstrated in the previous sections, the MRF method produces a larger quantity of successful velocity measurements with errors comparable to those produced with the SRF method. 
However, this does not guarantee that the flows determined by either method will be correct. 
To ascertain the accuracy of each fitting method, we must compare the velocity measurements to a known velocity. 
We accomplish this by introducing known velocities into the data set through the tile tracking rate. 
Usually every tile is tracked longitudinally at a rate close to the surface differential rotation rate in order to minimize the effects of a large zonal flow. 
By altering the tracking rate for each tile such that it slides east or west at some $\Delta v$ relative to the standard tracking rate, we introduce a known velocity as a Galilean transformation. 
We track the same set of tiles from the same temporal sequence at a variety of $\Delta v$ values and use both fitting procedures to gather velocity measurements. 
By comparing the difference in measured zonal velocity between a region tracked at $\Delta v = 0$ and the same region tracked at a non-zero $\Delta v$, we determine how accurately each fitting method can reproduce the known velocity. 
By considering only the differences in measured velocities made at the same disk position, any systematics that are independent of tracking rate are removed from the analysis.

\begin{figure}[]
	\centering
	\includegraphics{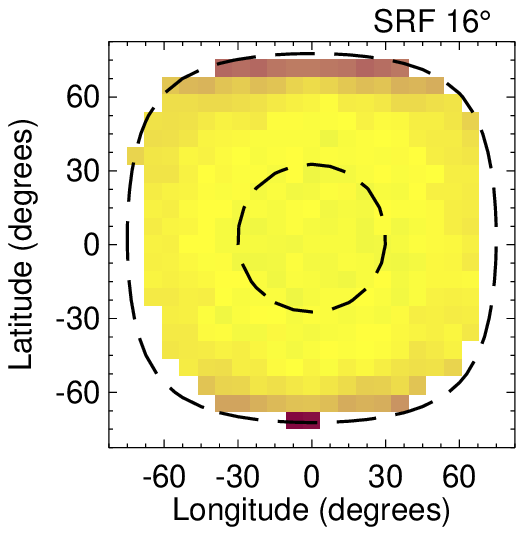}
	\includegraphics{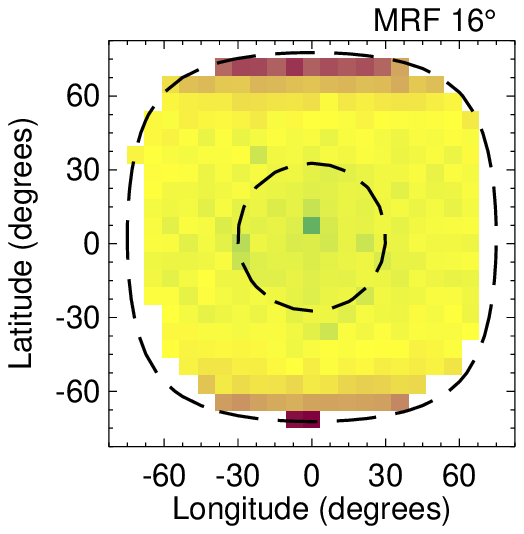}
	\includegraphics{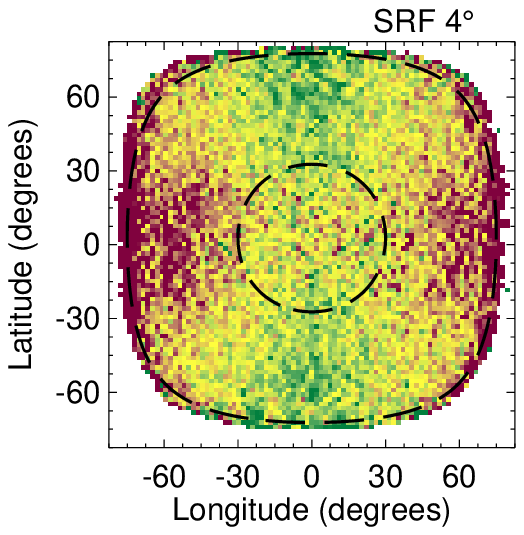}
	\includegraphics{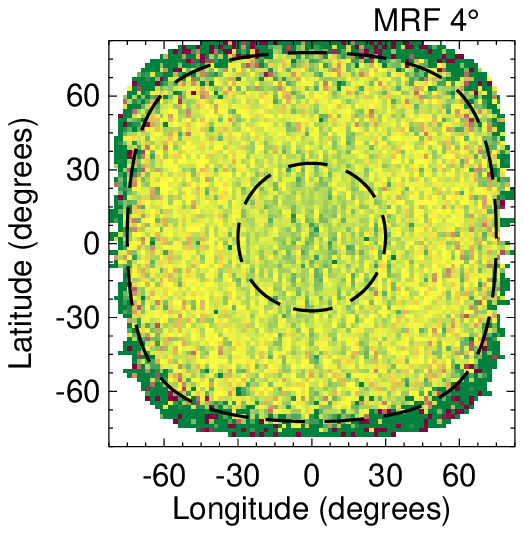}
	\includegraphics{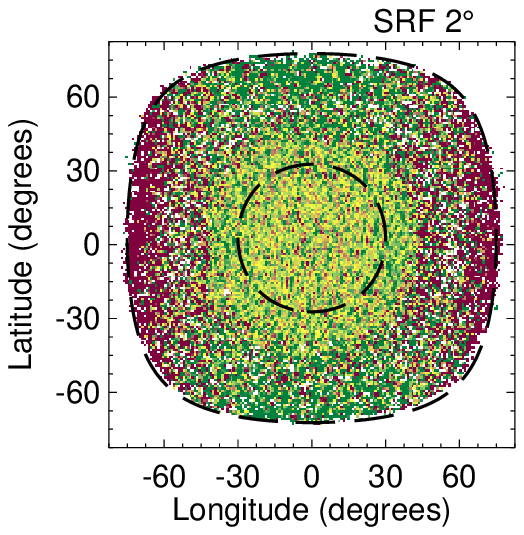}
	\includegraphics{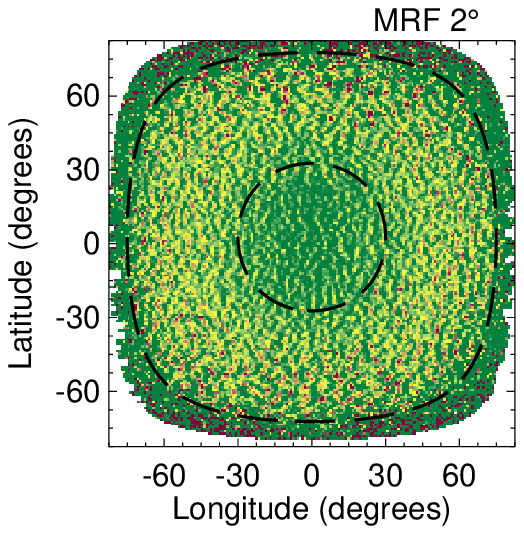}
	\\
	\includegraphics{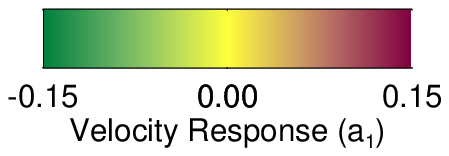}
	\caption{Velocity response for modes with lower turning points in the upper 16 Mm for $16^{\circ}$ tiles, 4 Mm for $4^{\circ}$ tiles, and 2 Mm for $2^{\circ}$ tiles. The inner dashed circle is at $30^{\circ}$ from disk center and the outer dashed line is at $75^{\circ}$. The SRF velocity response (left) has significantly higher spatial variability over the solar disk than the MRF method (right).}
	\label{fig:track4}
\end{figure}

\begin{figure}[]
	\centering
	\includegraphics{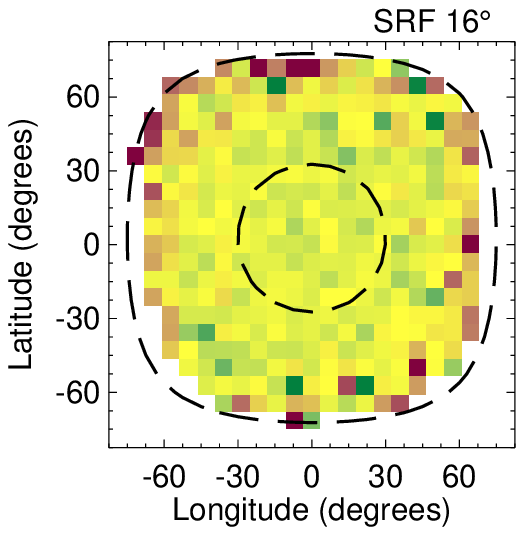}
	\includegraphics{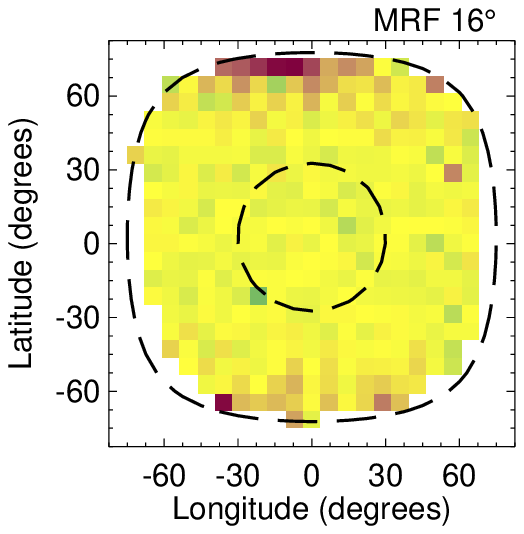}
	\includegraphics{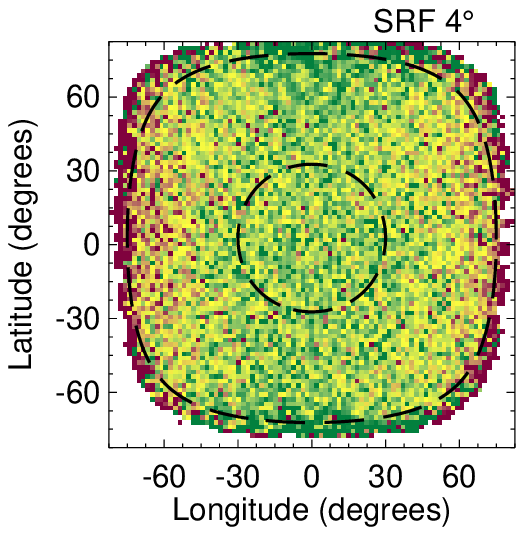}
	\includegraphics{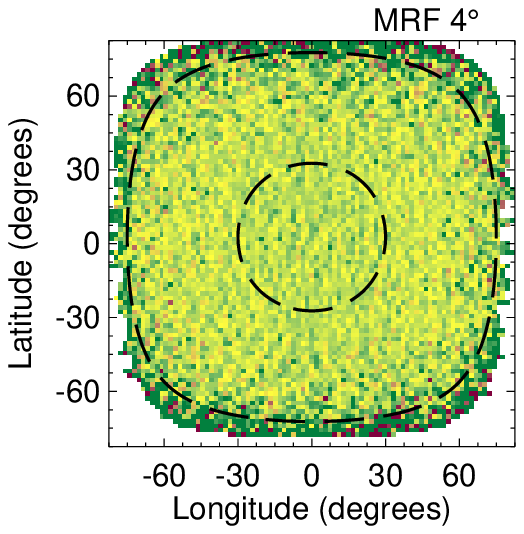}
	\includegraphics{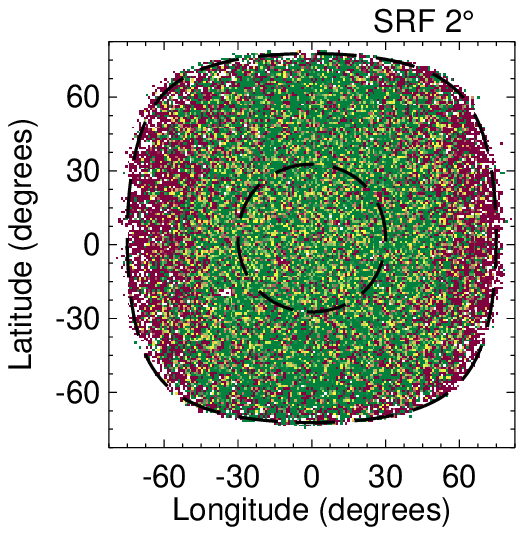}
	\includegraphics{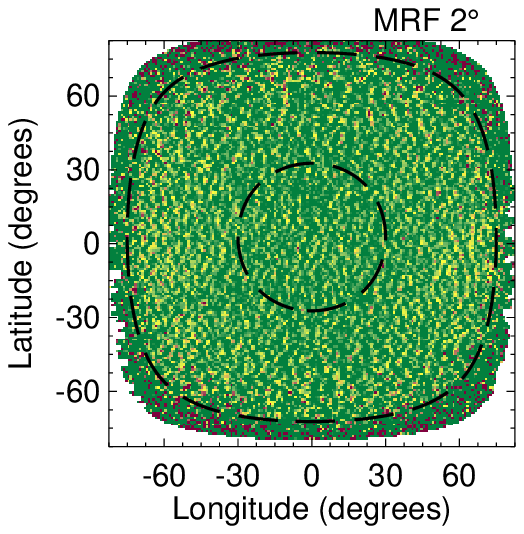}
	\\
	\includegraphics{track_map_colortable.ps}
	\caption{Velocity response for modes with lower turning points between 16 and 48 Mm for $16^{\circ}$ tiles, between 4 and 12 Mm for $4^{\circ}$, and between 2 and 6 Mm for $2^{\circ}$. The inner dashed circle is at $30^{\circ}$ from disk center and the outer dashed line is at $75^{\circ}$. Velocity measurements from deeper modes show less spatial variability over the solar disk for both the SRF method (left) and the MRF method (right).}
	\label{fig:track5}
\end{figure}

A set of tiles covering the entire solar disk are tracked for a single day over a range of $\Delta v$ spanning $500 \ \mathrm{m \ s^{-1}}$ in each direction.
The zonal-velocity measurements made at $\Delta v = 0$ are subtracted from those made at all other values of $\Delta v$ independently for each mode of each tile to remove physical flows and systematics. 
The ideal velocity response is to have the measured velocity difference consistent with the introduced velocity difference. 
By subtracting the known velocity difference from the measured difference, we obtain the velocity deviation as a function of $\Delta v$.
At each disk position, we perform a linear regression of these deviations against $\Delta v$ for a collection of modes of similar phase-speed. 

\begin{equation}
[u(\Delta v) - u_0] - \Delta v = a_0 + a_1 \Delta v
\label{eqn:delta}
\end{equation}
\begin{equation}
u_0 \equiv u(0)
\label{eqn:delta2}
\end{equation}
Here, $u(\Delta v)$ is the measured velocity of a single mode obtained with either fitting procedure for a given tile tracked at $\Delta v$ relative to the default rate, and $u_0$ is the measured velocity from the same tile tracked at the default rate. 
The left side of Equation \eqnref{eqn:delta} represents the deviation of measured velocity from the introduced velocity, while the right side expresses this deviation with a linear dependence on the introduced velocity. 
We bin together measurements made from modes within a range of phase-speeds at each disk position and fit the coefficients $a_0$ and $a_1$. 
We use the coefficient $a_1$ as a diagnostic for how each method responds to velocities where $a_1=0$ is ideal. 
Mapped as a function of disk position (Figures \ref{fig:track4}, \ref{fig:track5}), this quantity shows the spatial uniformity of velocity response. 
The measured value of $a_0$ is consistent with zero in all cases, so it will not be discussed further. 
The mode set for each tile size is divided into two subsets based on the lower turning-point depth of each mode. 
The depth at which each mode set is split is dependent on tile size and is chosen to get roughly the same number of modes in each subset. 
Modes with lower turning points in the upper 16 Mm for $16^{\circ}$ tiles, 4 Mm for $4^{\circ}$ tiles, and 2 Mm for $2^{\circ}$ tiles are shown in Figure \ref{fig:track4}, while modes with lower turning-point depths below this are shown in Figure \ref{fig:track5}.

The velocity response of modes with shallow lower turning points from $16^{\circ}$ tiles is largely ideal across the entire disk for both the SRF and MRF methods (Figure \ref{fig:track4}), but with a slight overestimation of velocities at high latitudes. 
The MRF 
For smaller tile sizes, the SRF method produces position-dependent results, while the MRF method begins to show a uniform underestimation of velocities. 
For $4^{\circ}$ tiles, the SRF method underestimates velocities far from the Equator by around $10\,\%$ and overestimates far from the central meridian by the same amount. 
The SRF pattern for $2^{\circ}$ tiles is much less clear due to scatter in the measurements and lower success rate, but there is still a trend for overestimating in the east--west direction and underestimating in the north--south direction. 
The MRF results exhibit a much more spatially uniform pattern while still showing a consistent underestimation of $5\,\%$ to $10\,\%$. 
For every tile size, the SRF results tend to be most accurate within $30^{\circ}$ of disk center (the inner dashed circles of Figures \ref{fig:track4} and \ref{fig:track5}) while the MRF results show a consistent underestimation that depends on tile size. 
From $30^{\circ}$ out to $75^{\circ}$ from disk center (the outer dashed lines of Figures \ref{fig:track4} and \ref{fig:track5}), the MRF results are more accurate and spatially consistent than the SRF results. 
Beyond $75^{\circ}$ for smaller tile sizes, the MRF results begin a consistent underestimation while the SRF results maintain the spatially varying pattern seen inside $75^{\circ}$.
For modes sensitive to deeper flows (Figure \ref{fig:track5}), spatial uniformity is less of an issue while both methods obtain underestimations ($5\,\%$ to $10\,\%$) of the introduced velocity.

As tiles are tracked at large values of $\Delta v$, they begin to average over the sub-surface flows in a more extended physical region than the same tiles tracked at $\Delta v = 0$. 
Since this extra tracking distance is constant for all tile sizes, it is the smallest tile sizes that are affected the most. 
Sub-surface flows that vary on a length scale comparable to the tile size add to the scatter seen in Figures \ref{fig:track4} and \ref{fig:track5}. 
Since the zonal component of these small-scale flows averages to zero over large regions of the disk, there is no significant bias introduced to the analysis. 
The small-scale pattern that emerges in the $2^{\circ}$ map for MRF measurements (and $4^{\circ}$ to a lesser extent) is due to supergranular flows. 
While the SRF maps should display the same exact pattern, it is interesting to note that the variance appears to obscure the pattern.

\section{Discussion and Conclusions}
\label{sec:discussion}

Extracting frequency shifts that are interpreted as sub-surface velocities is done through fitting a model of acoustic power to a three-dimensional power spectrum. 
By fitting a model that includes various effects seen in the data to different segments of the three-dimensional spectrum, many unique fitting methods can be created. 
We have presented two such fitting procedures that differ both in the model used and the selection of data considered in a single optimization. 
These methods use identical expressions for the velocity-induced frequency splitting for each oscillation mode, yet produce significantly different results for the velocity measurements themselves. 

\subsection{Improvements in Depth for Inversions}

The MRF method is able to obtain a greater number of velocity measurements for higher phase-speed modes in a given tile size than the SRF method. 
The increase in the success rate is also prominent for small tile sizes, suggesting that the treatment of mode blending plays an important role. 
Smaller tiles exhibit more leakage of modal power across wavenumbers than larger tiles due to spatial apodization. 
The spreading of power increases the effective width of modes at a single wavenumber. 
High-phase-speed modes are already spaced closer in wavenumber than other modes, enhancing the mode blending for their region of each spectrum. 
The SRF model does not accommodate any overlap of power from neighboring modes when fitting a single mode, causing a disparity between model and data that worsens at higher phase-speeds and smaller tile sizes. 
The MRF method, in accounting for mode overlap, returns a higher number of successful measurements for these cases that can be used for studying deep flows and high horizontal resolution flows. 

Referring to Equation \eqnref{eqn:kerneldef}, the set of measurements obtained from a single tile \noparen{$\bvec{u}_n(k)$} are measures of the true sub-surface velocity \noparen{$\bvec{v}(\bvec{r})$} within a single region \noparen{$d^3r$} weighted with different sensitivity kernels \noparen{$K(\bvec{r})$}.
The sensitivity kernels generally do not have sufficient isolation of sensitivity at any given depth, so while having a larger number of high-phase-speed measurements results in reaching deeper into the Sun, these measurements alone do not provide a localized determination of sub-surface flows. 
The velocity measurements from many modes can be merged through an inversion to produce an estimate of the true sub-surface flow velocity in an isolated region. 
After finding a linear combination of sensitivity kernels that produces a single isolated peak of sensitivity at a target location, the estimated velocity for that region is constructed using the same linear combination of the associated velocity measurements:
\begin{equation}
\bvec{w} = \sum_{i} a_{i} \bvec{u}_{i}.
\label{eqn:invsoln}
\end{equation}
\noindent
Here $\bvec{w}$ is the estimated sub-surface flow velocity at a selected target point, $\bvec{u_{i}}$ are the velocity measurements obtained from each mode $(n,k)$, and $a_{i}$ are the coefficients of the linear combination determined by the inversion. 
To estimate the localization of each solution point, we can construct an averaging kernel that represents the spatial distribution of sensitivity for a solution point, just as a sensitivity kernel does for measured velocities: 
\begin{equation}
H(\bvec{r}) = \sum_{i} a_{i} K_{i}(\bvec{r}).
\label{eqn:avgkercombination}
\end{equation}
The set of modes that we are able to fit reliably using either method determine the basis set of kernels that can be used to construct a locally isolated averaging kernel $H(\bvec{r})$. 
It is not only how high in phase-speed the fitting method can reach, but also how reliably it can obtain successful measurements at all phase-speeds, that determines how well we can constrain sub-surface flows. 

\begin{figure}[]
	\centering
	\includegraphics{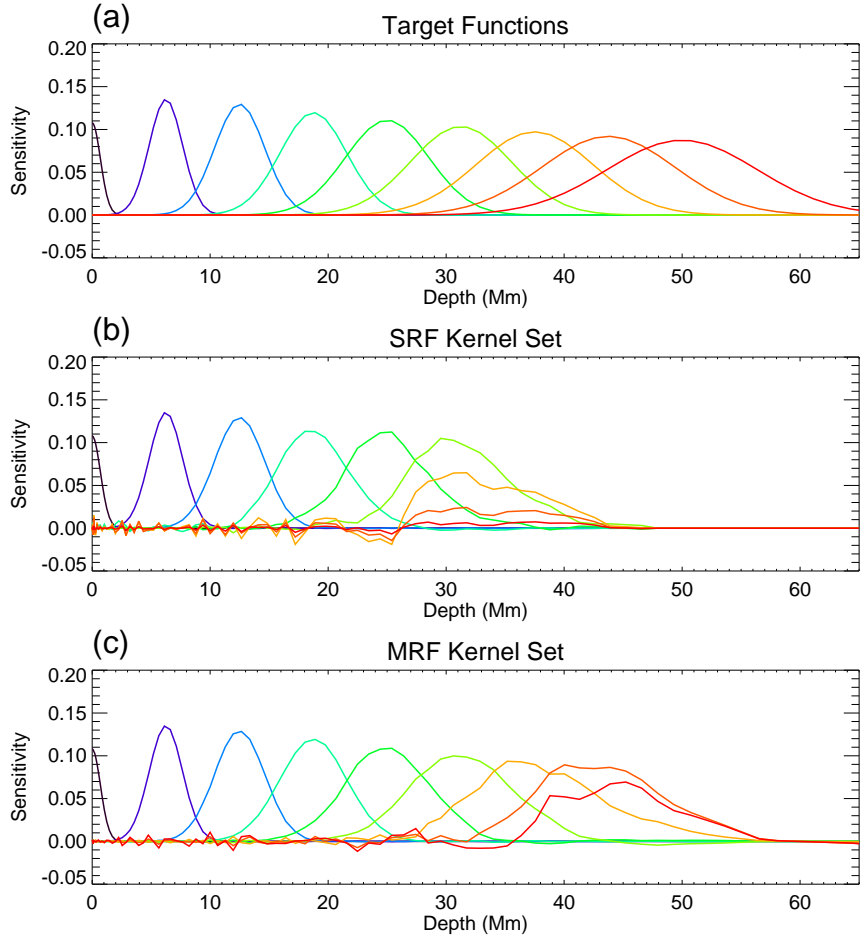}
	\caption{(a) Target functions for the 1D OLA inversion. 
	(b) Averaging kernels constructed from sensitivity kernels from the SRF mode set. 
	(c) Averaging kernels constructed from sensitivity kernels from the MRF mode set. 
	The colors indicate different target depths. 
	The SRF method is only able to produce averaging kernels down to a depth of about $30 \ \mathrm{Mm}$, while the MRF method reaches down to $45 \ \mathrm{Mm}$.}
	\label{fig:kernels}
\end{figure}

To determine how well averaging kernels can be constructed using the mode set available from each fitting method operating on $16^{\circ}$ tiles, we use a simple 1D optimally localized averaging (OLA) inversion (\opencite{dalsgaard_1990}; \opencite{schou_1998}). 
The target functions are Gaussian profiles in depth and we restrict the set of sensitivity kernels available for inversion to those with a $\mathrm{80\,\%}$ or higher success rate. 
Figure \ref{fig:kernels} compares averaging kernels constructed from each kernel set to the target functions used. 

For shallow targets, both sets are able to accurately recreate the Gaussian target function with minimal sidelobes in depth. 
As the target depth increases, the number of kernels available for constructing averaging kernels decreases. 
Past a target depth of 30 Mm, the SRF set does not contain enough kernels to create a sufficiently isolated peak of sensitivity. 
The MRF set still contains enough unique kernels that reach past this depth to construct reasonable averaging kernels down to 45 Mm.

\begin{figure}[]
	\centering
	\includegraphics{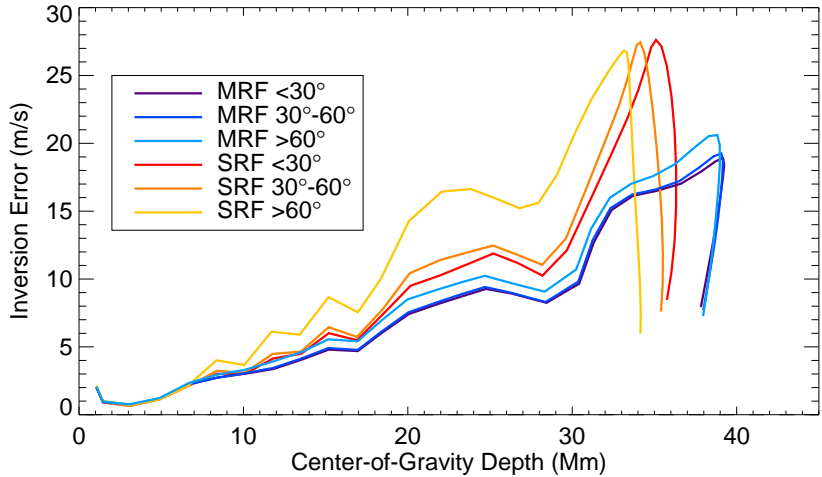}
	\caption{Typical error magnitudes propagated through 1D OLA inversion as a function of depth for multiple distances from disk center. 
	The sharp turnoff at the deep end of each curve demonstrates where proper averaging kernels can no longer be produced.
	The inversions done with SRF results tend to turn off between 33 and 35 Mm depth while those done with MRF results turn off between 38 and 40 Mm.}
	\label{fig:inverr}
\end{figure}

While the ability to construct reasonable averaging kernels through inversion is crucial for interpreting sub-surface flow results, this ignores the influence of measured errors. 
For OLA inversions, an optimization is done to balance the shape of the averaging kernel against the propagated error:
\begin{equation}
\sigma^2 = \sum_{i} ( a_{i} \sigma_{i} )^2.
\label{eqn:inverr}
\end{equation}
In this way, it is both the available mode set and the associated errors that determine the usefulness of the inversion results. 
Figure \ref{fig:inverr} shows the propagated inversion error as a function of the averaging kernel center-of-gravity depth for each fitting method. 
To mimic the result of inverting an average of many days of data over all longitudes to look at large-scale mean flows, the error for each mode has been scaled:
\begin{equation}
\sigma_i \rightarrow \frac{\sigma_{i}}{\sqrt{N_d N_l r_i}},
\label{eqn:inverr2}
\end{equation}
where $r_i$ is the success rate for a given mode, $N_d$ is the number of days to average over, and $N_l$ is the number of longitudes to average over. 
We set $N_d$ to 180 to mimic inversions of six months of data and $N_l$ to 17 for the typical number of tiles spanning the solar disk in longitude. 
The various curves for different regions of the solar disk demonstrate how much the dependence of the success rate and average error on disk position affect inversion results. 
The sharp downturn seen at the deep end of each curve indicates where the averaging kernels no longer resemble the target functions and are therefore useless.

Due to the higher success rate for each mode, the MRF method is able to produce useful inversion results to a greater depth than the SRF method. 
As the distance from disk center increases, the MRF results remain mostly constant while the SRF results start to deteriorate both in error magnitude and kernel depth. 
Despite larger average error in the MRF results for shallow modes seen in Figure \ref{fig:avgerr}, the higher success rate for these modes causes the inversion errors to not show a significant increase relative to the SRF results.

\subsection{Spatial Variability and Implications}

Anisotropy along the azimuthal direction of the power spectrum appears to be the primary cause of fitting problems near the solar limb. 
The SRF method tackles this issue by operating on spectra that have been ``flattened" along azimuth. 
This step not only amplifies the modal power in certain directions in order to eliminate anisotropy, but also amplifies the noise. 
By accounting for the natural variation of power along each ridge as well as in the background power, the MRF method is able to push closer to the solar limb. 
As seen in Figure \ref{fig:successdisk}, the mode set measureable by the SRF method is position-dependent. 
The higher-order modes see a drop in success rate closer to disk center than lower-order modes. 
The depth to which one can measure is then shallower near the limb. 
The MRF method, in contrast, allows for consistent determinations of sub-surface flows within $75^{\circ}$ of disk center.

Large tile sizes tend to provide the most accurate velocity measurements for both fitting methods, with a slight tendency for both to overestimate at high latitudes (Figure \ref{fig:track4}). 
Both methods exhibit spatial variability in the velocity response for smaller tile sizes, but only for low-phase-speed modes. 
The MRF response is radially symmetric with the worst velocity response occurring near disk center, while the worst regions of the SRF response appear far from disk center and have opposite signs between the north--south and east--west direction. 
While the magnitude of this variability is relatively small, there are significant implications when attempting to analyze small residual flows left over from the subtraction of large-scale flows. 
Tiles at high latitudes tracked at the Carrington rate see an overall flow speed of nearly $200 \ \mathrm{m \ s^{-1}}$ due to differential rotation. 
The systematic underestimation seen in the SRF method introduces an anomalous retrograde flow of around $20 \ \mathrm{m \ s^{-1}}$, similar to the magnitude of the center-to-limb velocity systematic for low-phase-speed modes \cite{greer_2013}. 
It is unclear what causes the systematic inaccuracy in either fitting method, although the strong dependence on disk position suggests a possible coupling between velocity measurements and power anisotropy. 

As mentioned in Section \ref{sec:mrf}, there is a significant increase in the computational cost when switching from the SRF method to the MRF method. 
While processing a day's worth of tiles with the SRF method is typically one of the fastest steps in the HMI Pipeline, the MRF method brings the process of fitting more in line with the time needed for tracking. 
It is important to consider when this additional computational burden is justified. 
The SRF method produces high-quality results for low-phase-speed modes and for large tiles near disk center. 
This leaves three distinct cases where the MRF method provides improvements: measuring higher-phase-speed modes, pushing closer to the limb, and using small tiles. 
The implications of this new procedure on determining sub-surface flows follow these three technical improvements. 
For a given tile size, we are able to extend our analysis deeper into the Sun while maintaining a constant horizontal and temporal resolution. 
Analysis to these extended depths can be performed consistently across most of the solar disk, providing uniform coverage over a larger fraction of the Sun. 
The increased reliability of small tile sizes permits higher horizontal resolution analysis of sub-surface flows. 

\textbf{Acknowledgements}
This work was supported by NASA through NASA grants NNX08AJ08G, NNX08AQ28G, and NNX09AB04G. 
The data used here are courtesy of NASA/SDO and the HMI science team. 
SDO is a NASA mission, and the HMI project is supported by NASA contract NAS5-02139.

\bibliographystyle{plain}
\bibliography{paper}		

\begin{thebibliography}{10}

\bibitem{anderson_1990}
E.~R. {Anderson}, T.~L. {Duvall}, Jr., and S.~M. {Jefferies}.
\newblock {Modeling of solar oscillation power spectra}.
\newblock {\em \apj}, 364:699--705, December 1990.

\bibitem{basu_1999}
S.~{Basu}, H.~M. {Antia}, and S.~C. {Tripathy}.
\newblock {Ring diagram analysis of near-surface flows in the sun}.
\newblock {\em \apj}, 512:458--470, February 1999.

\bibitem{birch_2007}
A.~C. {Birch}, L.~{Gizon}, B.~W. {Hindman}, and D.~A. {Haber}.
\newblock {The linear sensitivity of helioseismic ring diagrams to local
  flows}.
\newblock {\em \apj}, 662:730--737, June 2007.

\bibitem{bogart_2011a}
R.~S. {Bogart}, C.~{Baldner}, S.~{Basu}, D.~A. {Haber}, and M.~C.
  {Rabello-Soares}.
\newblock {HMI ring diagram analysis I. The processing pipeline}.
\newblock {\em J. Phys. Conf. Ser.}, 271(1):012008, January 2011.

\bibitem{dalsgaard_1996}
J.~{Christensen-Dalsgaard}, W.~{Dappen}, S.~V. {Ajukov}, E.~R. {Anderson},
  H.~M. {Antia}, S.~{Basu}, and {et al.}
\newblock {The current state of solar modeling}.
\newblock {\em Science}, 272:1286--1292, May 1996.

\bibitem{dalsgaard_1990}
J.~{Christensen-Dalsgaard}, J.~{Schou}, and M.~J. {Thompson}.
\newblock {A comparison of methods for inverting helioseismic data}.
\newblock {\em \mnras}, 242:353--369, February 1990.

\bibitem{gonzalezhernandez_2010}
I.~{Gonz{\'a}lez Hern{\'a}ndez}, R.~{Howe}, R.~{Komm}, and F.~{Hill}.
\newblock {Meridional circulation during the extended solar minimum: Another
  component of the torsional oscillation?}
\newblock {\em \apjl}, 713:L16--L20, April 2010.

\bibitem{greer_2013}
B.~{Greer}, B.~{Hindman}, and J.~{Toomre}.
\newblock {Center-to-Limb Velocity Systematic in Ring-Diagram Analysis}.
\newblock In K.~{Jain}, S.~C. {Tripathy}, F.~{Hill}, J.~W. {Leibacher}, and
  A.~A. {Pevtsov}, editors, {\em Astronomical Society of the Pacific Conference
  Series}, volume 478 of {\em Astronomical Society of the Pacific Conference
  Series}, page 199, December 2013.

\bibitem{haber_2002}
D.~A. {Haber}, B.~W. {Hindman}, J.~{Toomre}, R.~S. {Bogart}, R.~M. {Larsen},
  and F.~{Hill}.
\newblock {Evolving submerged meridional circulation cells within the upper
  convection zone revealed by ring-diagram analysis}.
\newblock {\em \apj}, 570:855--864, May 2002.

\bibitem{haber_2000}
D.~A. {Haber}, B.~W. {Hindman}, J.~{Toomre}, R.~S. {Bogart}, M.~J. {Thompson},
  and F.~{Hill}.
\newblock {Solar shear flows deduced from helioseismic dense-pack samplings of
  ring diagrams}.
\newblock {\em \solphys}, 192:335--350, March 2000.

\bibitem{harvey_1985}
J.~{Harvey}.
\newblock {High-resolution helioseismology}.
\newblock In E.~{Rolfe} and B.~{Battrick}, editors, {\em Future missions in
  solar, heliospheric \& space plasma physics}, volume 235 of {\em ESA Special
  Publication}, pages 199--208, June 1985.

\bibitem{hill_1988}
F.~{Hill}.
\newblock {Rings and trumpets - Three-dimensional power spectra of solar
  oscillations}.
\newblock {\em \apj}, 333:996--1013, October 1988.

\bibitem{hindman_2005}
B.~W. {Hindman}, D.~{Gough}, M.~J. {Thompson}, and J.~{Toomre}.
\newblock {Helioseismic ring analyses of artificial data computed for
  two-dimensional shearing flows}.
\newblock {\em \apj}, 621:512--523, March 2005.

\bibitem{komm_2013}
R.~{Komm}, I.~{Gonz{\'a}lez Hern{\'a}ndez}, F.~{Hill}, R.~{Bogart}, M.~C.
  {Rabello-Soares}, and D.~{Haber}.
\newblock {Subsurface meridional flow from HMI using the ring-diagram
  pipeline}.
\newblock {\em \solphys}, 287:85--106, October 2013.

\bibitem{upton_2012}
L.~{Rightmire-Upton}, D.~H. {Hathaway}, and K.~{Kosak}.
\newblock {Measurements of the sun's high-latitude meridional circulation}.
\newblock {\em \apjl}, 761:L14, December 2012.

\bibitem{routh_2011}
S.~{Routh}, D.~A. {Haber}, B.~W. {Hindman}, R.~S. {Bogart}, and J.~{Toomre}.
\newblock {The influence of tracking rate on helioseismic flow inferences}.
\newblock {\em J. Phys. Conf. Ser.}, 271(1):012014, January 2011.

\bibitem{scherrer_2012}
P.~H. {Scherrer}, J.~{Schou}, R.~I. {Bush}, A.~G. {Kosovichev}, R.~S. {Bogart},
  J.~T. {Hoeksema}, Y.~{Liu}, T.~L. {Duvall}, J.~{Zhao}, A.~M. {Title}, C.~J.
  {Schrijver}, T.~D. {Tarbell}, and S.~{Tomczyk}.
\newblock {The Helioseismic and Magnetic Imager (HMI) investigation for the
  Solar Dynamics Observatory (SDO)}.
\newblock {\em \solphys}, 275:207--227, January 2012.

\bibitem{schou_1998}
J.~{Schou}, H.~M. {Antia}, S.~{Basu}, R.~S. {Bogart}, R.~I. {Bush}, S.~M.
  {Chitre}, and {et al.}
\newblock {Helioseismic studies of differential rotation in the solar envelope
  by the solar oscillations investigation using the Michelson Doppler Imager}.
\newblock {\em \apj}, 505:390--417, September 1998.

\bibitem{snodgrass_1984}
H.~B. {Snodgrass}.
\newblock {Separation of large-scale photospheric Doppler patterns}.
\newblock {\em \solphys}, 94:13--31, August 1984.

\bibitem{zhao_2013}
J.~{Zhao}, R.~S. {Bogart}, A.~G. {Kosovichev}, T.~L. {Duvall}, Jr., and
  T.~{Hartlep}.
\newblock {Detection of equatorward meridional flow and evidence of double-cell
  meridional circulation inside the sun}.
\newblock {\em \apjl}, 774:L29, September 2013.

\bibitem{zhao_2012}
J.~{Zhao}, K.~{Nagashima}, R.~S. {Bogart}, A.~G. {Kosovichev}, and T.~L.
  {Duvall}, Jr.
\newblock {Systematic center-to-limb variation in measured helioseismic travel
  times and its effect on inferences of solar interior meridional flows}.
\newblock {\em \apjl}, 749:L5, April 2012.

\end{thebibliography}



\end{document}